\documentclass[a4paper,twocolumn,superscriptaddress,showpacs,floatfix,aps]{revtex4}

\usepackage{graphicx}
\usepackage{amsmath}
\usepackage{color}
\usepackage{times}

\newcommand{\Cnl}{C_{\mathrm{nl}}}

\begin{document}
\title{Critical temperature of a trapped Bose gas: comparison of theory and
experiment}
\author{Matthew J. Davis}
\email{mdavis@physics.uq.edu.au}

\affiliation{ ARC Centre of Excellence for Quantum-Atom Optics, School of
Physical Sciences, University of Queensland, Brisbane, QLD 
4072, Australia}
\author{P. Blair Blakie}
\affiliation{Department of Physics, University of Otago, P.O. Box 56, Dunedin,
New Zealand} 
\begin{abstract}

We apply the Projected Gross-Pitaevskii equation (PGPE) formalism to the
experimental problem of the shift in critical temperature 
$T_c$ of a harmonically confined Bose gas
as reported in Gerbier \emph{et al.} [Phys. Rev. Lett. \textbf{92}, 030405
(2004)].  The PGPE method includes critical fluctuations and we find the
results differ from various mean-field theories, and are in best agreement with
experimental data. To unequivocally observe beyond mean-field effects, however,
the experimental precision must either improve by an order of magnitude, or consider more
strongly interacting systems.  This is the
first application of a classical field method to make quantitative comparison
with experiment. 

\end{abstract} 

\maketitle

 The shift in critical temperature $T_c$ with interaction
strength for the homogeneous Bose gas has been the subject of numerous studies
spanning almost fifty years since the first calculations of Lee and Yang
\cite{Lee1957a,Lee1958a}.  While there is a  finite shift to the
chemical potential in mean-field (MF) theory, the shift of the critical temperature
is zero \cite{Baym2001a}. The leading order effect is due to long-wavelength  critical
fluctuations and is inherently non-perturbative. Using effective field
theory  it was determined that the shift is $\Delta T_c/T_c^0 = c a n^{1/3}$ 
where $n$ is the particle number density, $a$ is the s-wave scattering
length, and $c$ is a constant of order unity \cite{Baym1999a}.  Until recently
results for the value of $c$ disagreed by an order of magnitude and even sign,
as summarised in Fig.~1 of \cite{Arnold2001c}.  However, two calculations
performed using lattice Monte Carlo 
have settled the matter, and
confirm that the shift is in the positive direction with combined estimate of $c
\approx 1.31 \pm 0.02$ \cite{Kashurnikov2001a,Arnold2001c}.  A number of recent
improved results broadly agree, and useful discussions are provided
by Andersen \cite{Andersen2004a} and Holzmann \emph{et
al.}~\cite{Holzmann2004a}.

The
situation for the
harmonically confined Bose gas is somewhat different.  The ideal gas transition
temperature and de Broglie wavelength at $T_c$ are
\begin{eqnarray}
T_c^0 = \left(\frac{N}{\zeta(3)}\right)^{1/3}\frac{\hbar\bar{\omega}}{k_B},
\quad \lambda_0 = \left(\frac{2 \pi \hbar^2}{m k_B T_c^0}\right)^{1/2},
\end{eqnarray}
with $\bar{\omega} = (\omega_x \omega_y \omega_z)^{1/3}$.    There
is a shift in $T_c$ due to finite size effects \cite{Grossmann1995a} given by
$\Delta T_c/T_c^0 \simeq
-0.73 N^{-1/3} \bar{\omega}/\omega$
with $\omega =
(\omega_x + \omega_y + \omega_z)/3$, however this is usually small
for experimentally relevant parameters. 
The first-order shift in $T_c$ that survives in the thermodynamic limit is due
to mean-field effects and has been estimated analytically
\cite{Giorgini1996a}.  Repulsive interactions reduce $T_c$, which can be
intuitively understood due to a lowering of the peak density of the gas.
Next-order effects due to fluctuations have been
estimated in \cite{Houbiers1997a,Arnold2001a,Holzmann2004a} and in general 
predict an increase in $T_c$ from the first order result.
For a sufficiently wide trap Ref.~\cite{Arnold2001a} estimates
\begin{equation}
\frac{\Delta T_c}{T_c^0} = c_1\frac{a}{\lambda_0} + 
\left(c_2'\ln \frac{a}{\lambda_0} + c_2''\right)
\left(\frac{a}{\lambda_0}\right)^2 + O\left(\frac{a}{\lambda_0}\right)^3,
\label{eqn:2ndorder}
\end{equation}
with $c_1 = -3.426$, $c_2' = -45.86$, $c_2'' = -155.0$,
which for $a/\lambda_0 < 0.032$ predicts a positive shift due to fluctuations.
The first term is the MF result of \cite{Giorgini1996a}.
 Recently Zobay and co-authors have investigated  power law
traps with the goal of understanding how $T_c$ behaves in a smooth transition
from harmonic trapping to the homogeneous situation
\cite{Zobay2004a,Zobay2004b,Zobay2005a}. 

For a typical BEC experiment, the critical temperature deviates from the ideal
gas result only by a few percent.  As thermometry of Bose gases at this level
of accuracy can be difficult \cite{Gerbier2004b}, until recently the only
experimental measurement was reported by Ensher \emph{et al.}\  with $\Delta
T_c / T_c^0 = -0.06 \pm 0.05$ \cite{Ensher1996a}. However, in 2004
the Orsay group 
reported precise measurements of the critical temperature for a range
of atom numbers, and compared their results to the first-order MF estimate of
 \cite{Giorgini1996a}.  While in
agreement, the theoretical results lie near the upper range of the experimental
error bars. 


Previously one of us used the classical field projected Gross-Pitaevskii
equation (PGPE) formalism \cite{Davis2001a, Davis2001b, Davis2002a} to give an
estimate of the shift in $T_c$ of the homogeneous Bose gas \cite{Davis2003a}, which was found to
be in agreement with the Monte Carlo calculations \cite{Kashurnikov2001a,Arnold2001c}.  The PGPE
is a dynamical non-perturbative method, with the only approximation being that
the highly occupied modes ($\langle N_k \rangle \gg 1$) of the quantum Bose field  are
well-approximated by a classical field evolved according to the GPE.  
Related classical field approaches
have been considered by a number
of authors, including Kagan and co-workers \cite{Kagan1997c}, 
Sinatra \emph{et al.} \cite{Sinatra2000a,Sinatra2001a},
 Rz\c{a}\.{z}ewski and co-workers \cite{Goral2001a,Goral2002a}.

Here we use an extension of the PGPE for harmonically trapped gases
\cite{Blakie2005a} to calculate the shift in $T_c$ for the experiment of
Gerbier \emph{et al.}, and in particular focus on the competing effects of
mean-field and critical fluctuations.
The PGPE in dimensionless units is
\begin{eqnarray}
i\frac{\partial\Psi}{\partial \tau}  &=& 
-\nabla^{2}\Psi+V\Psi
 +
\mathcal{P}\{C_{\rm nl}|\Psi|^{2}\Psi\},
\label{eq:GPE1}
\end{eqnarray}
where $\Psi$ is the classical field, 
$V = (\lambda_x^{2}x^{2}+\lambda_y^{2}y^{2}+z^{2})/4$, and
$\lambda_{x,y} = \omega_{x,y}/\omega_z$.  The nonlinearity is 
$C_{\rm nl}=N_{b}U_{0}/\hbar\omega_{z}x_{0}^{3}$ where the unit of length is 
$x_{0}=\sqrt{\hbar/2m\omega_{z}}$ and $\tau = \omega_z t$.  
For the harmonic trap the Bose field is expanded on a basis of
harmonic oscillator eigenstates, with the cutoff energy $E_{\mathrm{cut}}$ 
determined by the occupation number condition.   The projection operator 
$\mathcal{P}\{F\}$
projects the function  $F$  onto the harmonic oscillator modes
with energy less than $E_{\mathrm{cut}}$.

The dynamical PGPE system represents a microcanonical ensemble, and  will evolve
any random initial conditions to thermal equilibrium defined by the
integrals of motion \cite{Davis2001b}.  For a cylindrically symmetric harmonic
trap these are the total number of particles, the energy, and the
component of the angular momentum along the symmetry axis.  Once in
equilibrium, we use the assumption of ergodicity to accurately
determine the condensate fraction \cite{Blakie2005a}, and the temperature $T$
and  chemical potential $\mu_{b}$ \cite{Davis2003a}.  By varying the initial
state energy  we measure the dependence of condensate fraction on
temperature.

\begin{figure}
\includegraphics[width=7cm]{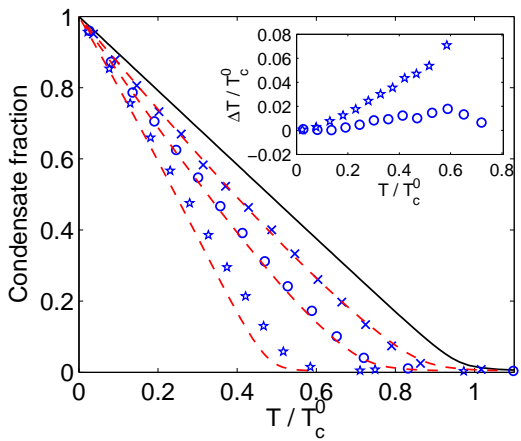}
\caption{(color online) Comparison of 
condensate fraction versus temperature normalised to $T_{c}^0$ for 
the PGPE method and the HFB-Popov calculation for
$\lambda_{x}=\sqrt{8},\,\lambda_{y}=1$ and $E_{{\rm cut}}=31\hbar\omega_{z}$.
   PGPE method:
$\Cnl = 500$ (crosses), $\Cnl = 2000$ (circles), $\Cnl = 10000$ (stars).
Dashed line: HFB-Popov results. Solid line: exact result for $\Cnl=0$.
Inset: $\Delta T = T_{\rm PGPE} - T_{\rm HFB}$ at
 fixed condensate fraction for  $\Cnl = 2000$ and $ 10000$, indicating a maximum shift near 
 $T_c$.
}
\label{fig:TOPresults}
\end{figure}

As an initial investigation into critical fluctuations, in
Fig.~\ref{fig:TOPresults} the results of the PGPE  calculations from
\cite{Blakie2005a} are compared with a self-consistent mean-field calculation
in the Popov approximation to the Hartree-Fock-Bogoliubov (HFB) formalism (see
e.g. \cite{Hutchinson1997a}). In order to make a direct comparison, the
HFB-Popov calculation is performed in the same basis as the dynamical PGPE
calculations, and we use the equipartition distribution $N_k = k_B
T/(\epsilon_k - \mu)$ for the quasi-particle occupations. (This is the high
temperature limit $k_B T \gg \epsilon_k$ of the Bose-Einstein distribution applicable to classical
fields).  For smaller values of $\Cnl$ the HFB-Popov theory agrees with the
classical field calculation, however for larger values there is a distinct
difference which we attribute to critical fluctuations.  We have repeated these
calculations using gapless implementations of HFB theory \cite{Proukakis1998a}
and found that they are little different from the results calculated using
HFB-Popov. Our results demonstrate that critical fluctuations have a measurable
effect for the PGPE system. However this is an idealised calculation --- to be
quantitative we must make a connection between the PGPE method and the recent
experiment \cite{Gerbier2004a}.

\begin{table}
\begin{tabular}{|l||c|c|c|c|c||c|c|c|c|c|}
\hline
$N_{\rm tot}^0$ ($10^6$) & 0.5 & 1.0 & 1.5 &  2.0 & 2.5 & 2.5 & 3.0 & 4.0 & 5.0 \\
$T_c^{0}$ (nK) &399 & 505 & 580 &  639 & 689 & 689 & 733 & 808 & 871 \\
$N_{\rm cut}$ &5.0 & 5.0 & 5.0 &  5.0 & 5.0 & 7.5 & 7.5 & 7.5 & 7.5\\
$E_{\rm cut}$ ($\hbar \omega_z$) & 219& 266 & 299 &  325 & 347 & 253 & 266 & 288& 307\\ 
Modes & 767 & 1382 & 1952 &   2498  & 3058 &  1172   &   1373  & 1730  &2129 \\
$N_{\rm b} (10^3)$ &8.75 & 15.0  & 20.7  &  26.1  & 31.4  & 19.2  & 22.1  & 27.6  & 33.1 \\
$\mu_c$ ($\hbar \omega_z$) & 101& 119 & 132 &  142 & 152 & 135 & 143 & 153& 163\\ 
$\delta_c $ ($\hbar \omega_z$) & 23& 29 & 34 & 37 & 41 & 39 & 41 & 46& 49\\ 
\hline
\end{tabular}
\caption{Input parameters for the PGPE simulations. The
chemical potential  $\mu_b$ and the shift of the cutoff energy 
$\delta_c$ are outputs parameters measured at the critical point.}
\label{simdata}
\end{table}

Gerbier \emph{et al.}\ \cite{Gerbier2004a} trap $^{87}$Rb atoms  in a
cylindrically symmetric harmonic potential with $(\omega_{x,y},\omega_z)  = 
2\pi \times (413,8.69)$ Hz giving $\lambda_{x,y} = 47.52$.  For total numbers
of atoms $N_{\mathrm{tot}}$ ranging from $2.5\times 10^5$ to $2.5 \times 10^6$,
the critical point was identified by reducing the final rf frequency of the
evaporative cooling, and identifying the point that the condensate fraction
became measurable (see Fig.~2 of  \cite{Gerbier2004a}.)  
We perform numerical simulations in a similar manner.  We choose relevant
simulation parameters and dynamically evolve the system to equilibrium for a
range of energies.  We identify the critical point from the number of
condensate particles and determine the number of particles above the cutoff
using a self-consistent semi-classical approximation for the high-energy modes
as described below.  This gives us a set of points $(N_c,
T_c)$ to be compared with the experimental data.

To simulate the experiments of Gerbier \emph{et al.}\ using the PGPE we need to choose both
an energy cutoff $E_{\rm cut}$ and a number of particles below the cutoff $N_b$
to simulate so that the occupation number condition is satisfied.  However, any
final result should be insensitive to the exact value of the cutoff that is
chosen.  \emph{A priori} estimates for our simulation parameters were determined from
the Bose-Einstein distribution of quantum orbitals of an ideal trapped gas at the
critical temperature, and are summarised in
Table~\ref{simdata}.
For the smaller clouds we chose
an energy cutoff such that $\langle N_k \rangle \ge 5$. 
For the large clouds this leads
to correspondingly larger basis sets that become computationally 
prohibitive, and for these we chose $\langle N_k \rangle \ge 7.5$.  
In principle we
could use this occupation condition for all simulations,
 however the two calculations at the crossover point 
 ($N_{\rm{tot}}^0=2.5\times10^6$) enables us to
verify that our calculations are insensitive to the exact value 
of the energy cutoff.  
We use the PGPE to evolve randomised initial states to
equilibrium and  measure the condensate number $N_0$, chemical potential 
$\mu_b$, temperature $T$, and density $n_b(\mathbf{x})$ for each set of
parameters.

\begin{figure}
\includegraphics[width=8.6cm]{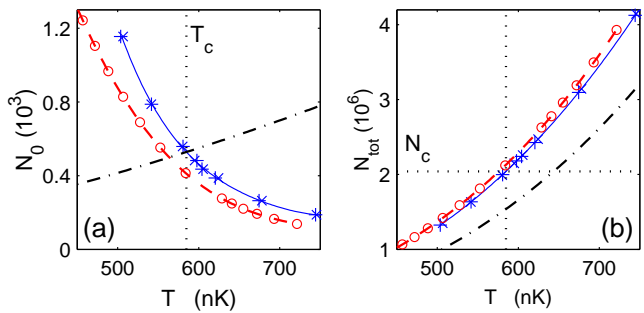}
\caption{(color online)
Determination of the critical number and temperature
 for the simulation with $N_{\mathrm{tot}}^0 = 4.0 \times 10^6$.
(a)
$N_0$ vs $T$ for: classical field (crosses, plusses, solid line),
HFB-Popov (circles, dashed line).  The number of condensate atoms for the ideal
gas at the critical temperature is the dot-dash line.
(b) Total number of atoms for: classical field (crosses, plusses, solid line),
HFB-Popov (circles, dashed line).  
Critical number versus temperature for the finite-sized ideal gas is 
the dot-dash line.
For both (a) and (b) the solid and dashed lines are polynomial
fits to the data.}
\label{fig:measureN}
\end{figure}

In Fig.~\ref{fig:measureN}(a) we plot the condensate number versus temperature
for the $N_{\rm tot}^0 = 4\times10^6$ data set and find
there is no sharp transition.  This is because we are only considering the atoms below the
cutoff. As the majority of atoms in the full system are above the cutoff and $N_0$ is of
order a few hundred particles for all the data points on this graph, these simulation
results all lie close to the critical point.  To determine a single critical point from
each data set we plot on the same graph the corresponding condensate number for the
finite-sized ideal gas at the same critical temperature.  
We choose the intersection of
these two curves to identify the critical point, and have verified that the occupation
number condition is satisfied here.

To relate these results back to the full experimental system we  assume that the
classical field and the above cutoff thermal cloud are weakly-coupled systems in
equilibrium, 
with the same temperature and chemical potential.  The
thermal cloud exists in the potential of the trap plus time-averaged classical
field density $n_b(\mathbf{x})$ that is determined from the PGPE simulations.
To solve for the above cutoff thermal cloud we make use of the self-consistent
Hartree-Fock approximation, which provides an accurate description of the modes
above $E_{\mathrm{cut}}$. The above cutoff density is determined by the
self-consistent solution of
\begin{equation}
n_{a}(\mathbf{x}) = 
\frac{1}{h^3} \int_{E_{\mathrm{HF}} > E_{0}} 
\hspace{-7mm} d^3 \mathbf{p} \left[
e^{(E_{\mathrm{HF}}(\mathbf{p},\mathbf{x}) - \mu)/k_B T} -1
\right]^{-1},
\label{eqn:scdensity}
\end{equation}
\begin{eqnarray}
E_{\mathrm{HF}}(\mathbf{p},\mathbf{x}) = 
p^2/2m + V_{\rm trap}(\mathbf{x}) + 
2 U_0 [n_{b}(\mathbf{x}) + n_{a}(\mathbf{x})],
\label{eqn:EHF}
\end{eqnarray}
where $E_{\mathrm{HF}}(\mathbf{p},\mathbf{x})$ is the Hartree-Fock energy. In
this procedure the contribution of the above cutoff density $n_a(\mathbf{x})$ 
to the effective potential for the classical field is neglected. This is
justified as we find that near the critical point $n_a(\mathbf{x})$ is approximately flat in
the region where the $n_b(\mathbf{x})$ is non-zero.   However, the uniform energy
shift of this interaction  must be included in the chemical potential used in
Eq.~(\ref{eqn:scdensity}) as  $\mu = \mu_b + 2 n_a(\mathbf{0}) U_0$. 
Another important correction accounts for the shift in the energy of
the highest  oscillator modes in the classical field from $E_{\mathrm{cut}}$ due
to interaction effects so that the integral in Eq.~(\ref{eqn:scdensity}) is over
the correct region of phase space.  We do this by assuming that the highest
energy modes of the classical field are single-particle in nature, and are
shifted by a constant amount $\delta_c$.
We fit the time-averaged occupation of
these modes to  $\langle N_k  \rangle = T/(\epsilon_k^0 + \delta_c - \mu_b)$. 
 The lower limit of
the integral in Eq.~(\ref{eqn:scdensity}) is then $ E_0 = E_{\mathrm{cut}}  +
\delta_c + 2 n_a(\mathbf{0}) U_0$ to account for the mean-field of
the thermal cloud.

\begin{figure}
\includegraphics[width=8.6cm]{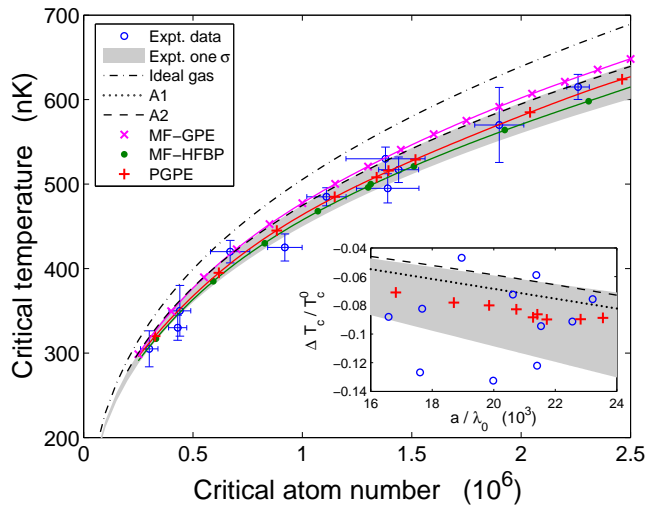}
\caption{(color online) Comparison of theoretical calculations with experiment.
The main figure plots $T_c$ vs $N_c$, whereas the inset plots
the shift of $T_c$ against the relevant small parameter $a/\lambda_0$. 
Experimental results: data  (open circles),  one $\sigma$ fit (grey area).
 Theoretical
results for $T_c$: ideal gas (dot-dash), A1 (dotted), A2 (dashed),
MF-GPE (crosses), MF-HFBP
 (dots),
PGPE (plusses).  Solid lines through the data points are polynomial fits. 
A1 is not shown in the main figure for clarity.  
}
\label{fig:exptcomparison}
\end{figure}

We have also calculated $T_c$ using other methods for comparison, as
summarised below:\\
\noindent 1. A1: 
This is the first order analytic estimate of Giorgini 
\emph{et al.}~\cite{Giorgini1996a}, which is the first term of 
Eq.~(\ref{eqn:2ndorder}).\\
\noindent 2. A2:  This is the full second order result of
Eq.~(\ref{eqn:2ndorder}).  However, the validity condition for
 this result (Eq.~(7.2) of \cite{Arnold2001a})
requires the trap to be ``sufficiently broad'', and this 
is strongly violated for this experiment.  This
essentially says that the semi-classical
approximation is not valid for the lowest energy modes of this 
strongly elongated system.\\
\noindent 3. MF-GPE: 
The GPE is solved
numerically using a variational Gaussian ansatz, and the thermal cloud
calculated using a semi-classical approximation \cite{Giorgini1996a}.  At each temperature the
condensate and non-condensate are determined self-consistently with a fixed
number of particles, and the critical
temperature is where the condensate fraction decreases to zero.\\
\noindent 4. MF-HFBP: We fix the condensate fraction, and determine the
temperature that gives an appropriate self-consistent condensate mode and
thermal density.  We have verified the results are unchanged for
equipartition or Bose-Einstein statistics.  We
use the same procedure as for the PGPE calculation to determine the
critical point, the above cutoff density and the total atom number.  An
illustrative set of data is displayed in Fig.~\ref{fig:measureN}.

In Fig.~\ref{fig:exptcomparison} we compare these theoretical results with
the PGPE and experimental data.  The MF A1 estimate
was shown in
\cite{Gerbier2004a} and is within the experimental error bars.   However, our
more accurate MF-GPE calculation gives a greater value of
$T_c$ at larger atom numbers, agreeing with 
the mean-field results of 
Houbiers
\emph{et al.}~\cite{Houbiers1997a}. 
However, the MF-HFBP result, which presumably is an even better mean-field
calculation, is quite
different and towards the lower end of the experimental error estimate.

The predicted effect of critical fluctuations \cite{Houbiers1997a,Arnold2001a}
is to \emph{further} increase $T_c$.  The non-perturbative A2 estimate lies at
the boundary of experimental error, but as mentioned earlier this result does
not satisfy the validity requirement for this experiment. The PGPE calculation,
which includes all the physics of the MF-HFBP calculation as well as critical
fluctuations, is measurably different.  Arguably it is in best
agreement with the experimental data.  However, both the PGPE and  MF-HFBP
calculations lie within the error bars, suggesting that experimental precision
must improve by an order of magnitude in order to distinguish these
predictions.

The inset of  Fig.~\ref{fig:exptcomparison} shows the PGPE shift as a function
of $a/\lambda_0$ and in comparison with the results
of Eq.~(\ref{eqn:2ndorder}) and the experimental data.  The second order term is almost constant over the
experimental range of $a/\lambda_0$ and so cannot distinguish the presence or
otherwise of any logarithmic term.  We note that the finite-size shift is
subtracted from the PGPE and experimental data for this comparison.

We have also translated data for parameters as in Fig.~\ref{fig:TOPresults}
but with $\Cnl=5000$ to
realistic experimental values, and found that for $10^7$  atoms of $^{87}$Rb in
a TOP trap with a 40 Hz radial frequency that
the difference between the MF-HFBP and PGPE results is of order 3\%.  Thus we
suggest that for currently accessible experimental conditions it will be
necessary to either make use of Feshbach resonances to probe more strongly interacting
regimes, or to move to traps flatter than harmonic to be able to
distinguish these theories in the lab.

In conclusion we have performed a careful theoretical analysis of the
experiment on the shift in critical temperature of a trapped Bose gas reported
in Gerbier \emph{et al.}~\cite{Gerbier2004a}.  We have determined that earlier
calculations based on mean-field theory and the local-density approximation are
inappropriate for this experiment, and make predications for $T_c$ outside the
experimental error bars at larger atom numbers. We have applied
non-perturbative classical field theory to this problem, and described how to
incorporate the physics of the above-cutoff atoms in equilibrium.  The results
include the effect of critical fluctuations,  and give the best agreement with
experimental observations.  Our results indicate the precision
requirements for experiments to investigate beyond mean-field effects on the
critical temperature.

We thank Alain Aspect for useful discussions.  MJD
acknowledges financial support from the Australian Research Council and the
University of Queensland, and 
PBB from the Marsden Fund of New Zealand and the University of Otago.


\begin{thebibliography}{30}
\expandafter\ifx\csname natexlab\endcsname\relax\def\natexlab#1{#1}\fi
\expandafter\ifx\csname bibnamefont\endcsname\relax
  \def\bibnamefont#1{#1}\fi
\expandafter\ifx\csname bibfnamefont\endcsname\relax
  \def\bibfnamefont#1{#1}\fi
\expandafter\ifx\csname citenamefont\endcsname\relax
  \def\citenamefont#1{#1}\fi
\expandafter\ifx\csname url\endcsname\relax
  \def\url#1{\texttt{#1}}\fi
\expandafter\ifx\csname urlprefix\endcsname\relax\def\urlprefix{URL }\fi
\providecommand{\bibinfo}[2]{#2}
\providecommand{\eprint}[2][]{\url{#2}}

\bibitem[{\citenamefont{Lee and Yang}(1957)}]{Lee1957a}
\bibinfo{author}{\bibfnamefont{T.~D.} \bibnamefont{Lee}} \bibnamefont{and}
  \bibinfo{author}{\bibfnamefont{C.~N.} \bibnamefont{Yang}},
  \bibinfo{journal}{Phys. Rev.} \textbf{\bibinfo{volume}{105}},
  \bibinfo{pages}{1119} (\bibinfo{year}{1957}).

\bibitem[{\citenamefont{Lee and Yang}(1958)}]{Lee1958a}
\bibinfo{author}{\bibfnamefont{T.~D.} \bibnamefont{Lee}} \bibnamefont{and}
  \bibinfo{author}{\bibfnamefont{C.~N.} \bibnamefont{Yang}},
  \bibinfo{journal}{Phys. Rev.} \textbf{\bibinfo{volume}{112}},
  \bibinfo{pages}{1419} (\bibinfo{year}{1958}).

\bibitem[{\citenamefont{Baym et~al.}(2001)\citenamefont{Baym, Blaizot,
  Holzmann, Lalo{\"{e}}, and Vautherin}}]{Baym2001a}
\bibinfo{author}{\bibfnamefont{G.}~\bibnamefont{Baym}},
  \bibinfo{author}{\bibfnamefont{J.-P.} \bibnamefont{Blaizot}},
  \bibinfo{author}{\bibfnamefont{M.}~\bibnamefont{Holzmann}},
  \bibinfo{author}{\bibfnamefont{F.}~\bibnamefont{Lalo{\"{e}}}},
  \bibnamefont{and}
  \bibinfo{author}{\bibfnamefont{D.}~\bibnamefont{Vautherin}},
  \bibinfo{journal}{Eur. Phys. J. B} \textbf{\bibinfo{volume}{24}},
  \bibinfo{pages}{107} (\bibinfo{year}{2001}).

\bibitem[{\citenamefont{Baym et~al.}(1999)\citenamefont{Baym, Blaizot,
  Holzmann, Lalo{\"{e}}, and Vautherin}}]{Baym1999a}
\bibinfo{author}{\bibfnamefont{G.}~\bibnamefont{Baym}},
  \bibinfo{author}{\bibfnamefont{J.-P.} \bibnamefont{Blaizot}},
  \bibinfo{author}{\bibfnamefont{M.}~\bibnamefont{Holzmann}},
  \bibinfo{author}{\bibfnamefont{F.}~\bibnamefont{Lalo{\"{e}}}},
  \bibnamefont{and}
  \bibinfo{author}{\bibfnamefont{D.}~\bibnamefont{Vautherin}},
  \bibinfo{journal}{Phys. Rev. Lett.} \textbf{\bibinfo{volume}{83}},
  \bibinfo{pages}{1703} (\bibinfo{year}{1999}).

\bibitem[{\citenamefont{Arnold and Moore}(2001)}]{Arnold2001c}
\bibinfo{author}{\bibfnamefont{P.}~\bibnamefont{Arnold}} \bibnamefont{and}
  \bibinfo{author}{\bibfnamefont{G.}~\bibnamefont{Moore}},
  \bibinfo{journal}{Phys. Rev. Lett.} \textbf{\bibinfo{volume}{87}},
  \bibinfo{pages}{120401} (\bibinfo{year}{2001}).

\bibitem[{\citenamefont{Kashurnikov et~al.}(2001)\citenamefont{Kashurnikov,
  Prokof'ev, and Svistunov}}]{Kashurnikov2001a}
\bibinfo{author}{\bibfnamefont{V.~A.} \bibnamefont{Kashurnikov}},
  \bibinfo{author}{\bibfnamefont{N.~V.} \bibnamefont{Prokof'ev}},
  \bibnamefont{and} \bibinfo{author}{\bibfnamefont{B.~V.}
  \bibnamefont{Svistunov}}, \bibinfo{journal}{Phys. Rev. Lett.}
  \textbf{\bibinfo{volume}{87}}, \bibinfo{pages}{120402}
  (\bibinfo{year}{2001}).

\bibitem[{\citenamefont{Andersen}(2004)}]{Andersen2004a}
\bibinfo{author}{\bibfnamefont{J.~O.} \bibnamefont{Andersen}},
  \bibinfo{journal}{Rev. Mod. Phys.} \textbf{\bibinfo{volume}{76}},
  \bibinfo{pages}{599} (\bibinfo{year}{2004}).

\bibitem[{\citenamefont{Holzmann et~al.}(2004)\citenamefont{Holzmann, Fuchs,
  Baym, Blaizot, and Lalo{\"e}}}]{Holzmann2004a}
\bibinfo{author}{\bibfnamefont{M.}~\bibnamefont{Holzmann}},
  \bibinfo{author}{\bibfnamefont{J.-N.} \bibnamefont{Fuchs}},
  \bibinfo{author}{\bibfnamefont{G.~A.} \bibnamefont{Baym}},
  \bibinfo{author}{\bibfnamefont{J.-P.} \bibnamefont{Blaizot}},
  \bibnamefont{and}
  \bibinfo{author}{\bibfnamefont{F.}~\bibnamefont{Lalo{\"e}}},
  \bibinfo{journal}{C. R. Physique} \textbf{\bibinfo{volume}{5}},
  \bibinfo{pages}{21} (\bibinfo{year}{2004}).

\bibitem[{\citenamefont{Grossmann and Holthaus}(1995)}]{Grossmann1995a}
\bibinfo{author}{\bibfnamefont{S.}~\bibnamefont{Grossmann}} \bibnamefont{and}
  \bibinfo{author}{\bibfnamefont{M.}~\bibnamefont{Holthaus}},
  \bibinfo{journal}{Phys. Lett. A} \textbf{\bibinfo{volume}{208}},
  \bibinfo{pages}{188} (\bibinfo{year}{1995}).

\bibitem[{\citenamefont{Giorgini et~al.}(1996)\citenamefont{Giorgini,
  Pitaevskii, and Stringari}}]{Giorgini1996a}
\bibinfo{author}{\bibfnamefont{S.}~\bibnamefont{Giorgini}},
  \bibinfo{author}{\bibfnamefont{L.~P.} \bibnamefont{Pitaevskii}},
  \bibnamefont{and}
  \bibinfo{author}{\bibfnamefont{S.}~\bibnamefont{Stringari}},
  \bibinfo{journal}{Phys. Rev. A} \textbf{\bibinfo{volume}{54}},
  \bibinfo{pages}{R4633} (\bibinfo{year}{1996}).

\bibitem[{\citenamefont{Houbiers et~al.}(1997)\citenamefont{Houbiers, Stoof,
  and Cornell}}]{Houbiers1997a}
\bibinfo{author}{\bibfnamefont{M.}~\bibnamefont{Houbiers}},
  \bibinfo{author}{\bibfnamefont{H.~T.~C.} \bibnamefont{Stoof}},
  \bibnamefont{and} \bibinfo{author}{\bibfnamefont{E.~A.}
  \bibnamefont{Cornell}}, \bibinfo{journal}{Phys. Rev. A}
  \textbf{\bibinfo{volume}{56}}, \bibinfo{pages}{2041} (\bibinfo{year}{1997}).

\bibitem[{\citenamefont{Arnold and Tom{\'a}\v{s}ik}(2001)}]{Arnold2001a}
\bibinfo{author}{\bibfnamefont{P.}~\bibnamefont{Arnold}} \bibnamefont{and}
  \bibinfo{author}{\bibfnamefont{B.}~\bibnamefont{Tom{\'a}\v{s}ik}},
  \bibinfo{journal}{Phys. Rev. A} \textbf{\bibinfo{volume}{64}},
  \bibinfo{pages}{053609} (\bibinfo{year}{2001}).

\bibitem[{\citenamefont{Zobay}(2004)}]{Zobay2004a}
\bibinfo{author}{\bibfnamefont{O.}~\bibnamefont{Zobay}}, \bibinfo{journal}{J.
  Phys. B} \textbf{\bibinfo{volume}{37}}, \bibinfo{pages}{2593}
  (\bibinfo{year}{2004}).

\bibitem[{\citenamefont{Zobay et~al.}(2004)\citenamefont{Zobay, Metikas, and
  Alber}}]{Zobay2004b}
\bibinfo{author}{\bibfnamefont{O.}~\bibnamefont{Zobay}},
  \bibinfo{author}{\bibfnamefont{G.}~\bibnamefont{Metikas}}, \bibnamefont{and}
  \bibinfo{author}{\bibfnamefont{G.}~\bibnamefont{Alber}},
  \bibinfo{journal}{Phys. Rev. A} \textbf{\bibinfo{volume}{69}},
  \bibinfo{pages}{063615} (\bibinfo{year}{2004}).

\bibitem[{\citenamefont{Zobay et~al.}(2005)\citenamefont{Zobay, Metikas, and
  Kleinert}}]{Zobay2005a}
\bibinfo{author}{\bibfnamefont{O.}~\bibnamefont{Zobay}},
  \bibinfo{author}{\bibfnamefont{G.}~\bibnamefont{Metikas}}, \bibnamefont{and}
  \bibinfo{author}{\bibfnamefont{H.}~\bibnamefont{Kleinert}},
  \bibinfo{journal}{Phys. Rev. A} \textbf{\bibinfo{volume}{71}},
  \bibinfo{pages}{043614} (\bibinfo{year}{2005}).

\bibitem[{\citenamefont{Gerbier
  et~al.}(2004{\natexlab{a}})\citenamefont{Gerbier, Thywissen, Richard,
  Hugbart, Bouyer, and Aspect}}]{Gerbier2004b}
\bibinfo{author}{\bibfnamefont{F.}~\bibnamefont{Gerbier}},
  \bibinfo{author}{\bibfnamefont{J.~H.} \bibnamefont{Thywissen}},
  \bibinfo{author}{\bibfnamefont{S.}~\bibnamefont{Richard}},
  \bibinfo{author}{\bibfnamefont{M.}~\bibnamefont{Hugbart}},
  \bibinfo{author}{\bibfnamefont{P.}~\bibnamefont{Bouyer}}, \bibnamefont{and}
  \bibinfo{author}{\bibfnamefont{A.}~\bibnamefont{Aspect}},
  \bibinfo{journal}{Phys. Rev. A} \textbf{\bibinfo{volume}{70}},
  \bibinfo{pages}{013607} (\bibinfo{year}{2004}{\natexlab{a}}).

\bibitem[{\citenamefont{Ensher et~al.}(1996)\citenamefont{Ensher, Jin,
  Matthews, Wieman, and Cornell}}]{Ensher1996a}
\bibinfo{author}{\bibfnamefont{J.~R.} \bibnamefont{Ensher}},
  \bibinfo{author}{\bibfnamefont{D.~S.} \bibnamefont{Jin}},
  \bibinfo{author}{\bibfnamefont{M.~R.} \bibnamefont{Matthews}},
  \bibinfo{author}{\bibfnamefont{C.~E.} \bibnamefont{Wieman}},
  \bibnamefont{and} \bibinfo{author}{\bibfnamefont{E.~A.}
  \bibnamefont{Cornell}}, \bibinfo{journal}{Phys. Rev. Lett.}
  \textbf{\bibinfo{volume}{77}}, \bibinfo{pages}{4984} (\bibinfo{year}{1996}).

\bibitem[{\citenamefont{Davis et~al.}(2001{\natexlab{a}})\citenamefont{Davis,
  Ballagh, and Burnett}}]{Davis2001a}
\bibinfo{author}{\bibfnamefont{M.~J.} \bibnamefont{Davis}},
  \bibinfo{author}{\bibfnamefont{R.~J.} \bibnamefont{Ballagh}},
  \bibnamefont{and} \bibinfo{author}{\bibfnamefont{K.}~\bibnamefont{Burnett}},
  \bibinfo{journal}{J. Phys. B} \textbf{\bibinfo{volume}{34}},
  \bibinfo{pages}{4487} (\bibinfo{year}{2001}{\natexlab{a}}).

\bibitem[{\citenamefont{Davis et~al.}(2001{\natexlab{b}})\citenamefont{Davis,
  Morgan, and Burnett}}]{Davis2001b}
\bibinfo{author}{\bibfnamefont{M.~J.} \bibnamefont{Davis}},
  \bibinfo{author}{\bibfnamefont{S.~A.} \bibnamefont{Morgan}},
  \bibnamefont{and} \bibinfo{author}{\bibfnamefont{K.}~\bibnamefont{Burnett}},
  \bibinfo{journal}{Phys. Rev. Lett.} \textbf{\bibinfo{volume}{87}},
  \bibinfo{pages}{160402} (\bibinfo{year}{2001}{\natexlab{b}}).

\bibitem[{\citenamefont{Davis et~al.}(2002)\citenamefont{Davis, Morgan, and
  Burnett}}]{Davis2002a}
\bibinfo{author}{\bibfnamefont{M.~J.} \bibnamefont{Davis}},
  \bibinfo{author}{\bibfnamefont{S.~A.} \bibnamefont{Morgan}},
  \bibnamefont{and} \bibinfo{author}{\bibfnamefont{K.}~\bibnamefont{Burnett}},
  \bibinfo{journal}{Phys. Rev. A.} \textbf{\bibinfo{volume}{66}},
  \bibinfo{pages}{053618} (\bibinfo{year}{2002}).

\bibitem[{\citenamefont{Davis and Morgan}(2003)}]{Davis2003a}
\bibinfo{author}{\bibfnamefont{M.~J.} \bibnamefont{Davis}} \bibnamefont{and}
  \bibinfo{author}{\bibfnamefont{S.~A.} \bibnamefont{Morgan}},
  \bibinfo{journal}{Phys. Rev. A} \textbf{\bibinfo{volume}{68}},
  \bibinfo{pages}{053615} (\bibinfo{year}{2003}).

\bibitem[{\citenamefont{Kagan and Svistunov}(1997)}]{Kagan1997c}
\bibinfo{author}{\bibfnamefont{Y.}~\bibnamefont{Kagan}} \bibnamefont{and}
  \bibinfo{author}{\bibfnamefont{B.~V.} \bibnamefont{Svistunov}},
  \bibinfo{journal}{Phys. Rev. Lett.} \textbf{\bibinfo{volume}{79}},
  \bibinfo{pages}{3331} (\bibinfo{year}{1997}).

\bibitem[{\citenamefont{Sinatra et~al.}(2000)\citenamefont{Sinatra, Castin, and
  Lobo}}]{Sinatra2000a}
\bibinfo{author}{\bibfnamefont{A.}~\bibnamefont{Sinatra}},
  \bibinfo{author}{\bibfnamefont{Y.}~\bibnamefont{Castin}}, \bibnamefont{and}
  \bibinfo{author}{\bibfnamefont{C.}~\bibnamefont{Lobo}}, \bibinfo{journal}{J.
  Mod. Opt.} \textbf{\bibinfo{volume}{47}}, \bibinfo{pages}{2629}
  (\bibinfo{year}{2000}).

\bibitem[{\citenamefont{Sinatra et~al.}(2001)\citenamefont{Sinatra, Lobo, and
  Castin}}]{Sinatra2001a}
\bibinfo{author}{\bibfnamefont{A.}~\bibnamefont{Sinatra}},
  \bibinfo{author}{\bibfnamefont{C.}~\bibnamefont{Lobo}}, \bibnamefont{and}
  \bibinfo{author}{\bibfnamefont{Y.}~\bibnamefont{Castin}},
  \bibinfo{journal}{Phys. Rev. Lett.} \textbf{\bibinfo{volume}{87}},
  \bibinfo{pages}{210404} (\bibinfo{year}{2001}).

\bibitem[{\citenamefont{G{\'o}ral et~al.}(2001)\citenamefont{G{\'o}ral,
  Englert, and Rz\c{a}\.{z}ewski}}]{Goral2001a}
\bibinfo{author}{\bibfnamefont{K.}~\bibnamefont{G{\'o}ral}},
  \bibinfo{author}{\bibfnamefont{B.-G.} \bibnamefont{Englert}},
  \bibnamefont{and}
  \bibinfo{author}{\bibfnamefont{K.}~\bibnamefont{Rz\c{a}\.{z}ewski}},
  \bibinfo{journal}{Phys. Rev. A} \textbf{\bibinfo{volume}{63}},
  \bibinfo{pages}{033606} (\bibinfo{year}{2001}).

\bibitem[{\citenamefont{G{\'o}ral et~al.}(2002)\citenamefont{G{\'o}ral, Gajda,
  and Rz\c{a}\.{z}ewski}}]{Goral2002a}
\bibinfo{author}{\bibfnamefont{K.}~\bibnamefont{G{\'o}ral}},
  \bibinfo{author}{\bibfnamefont{M.}~\bibnamefont{Gajda}}, \bibnamefont{and}
  \bibinfo{author}{\bibfnamefont{K.}~\bibnamefont{Rz\c{a}\.{z}ewski}},
  \bibinfo{journal}{Phys. Rev. A} \textbf{\bibinfo{volume}{66}},
  \bibinfo{pages}{051602(R)} (\bibinfo{year}{2002}).

\bibitem[{\citenamefont{Blakie and Davis}(2005)}]{Blakie2005a}
\bibinfo{author}{\bibfnamefont{P.~B.} \bibnamefont{Blakie}} \bibnamefont{and}
  \bibinfo{author}{\bibfnamefont{M.~J.} \bibnamefont{Davis}},
  \bibinfo{journal}{Phys. Rev. A} \textbf{\bibinfo{volume}{72}},
  \bibinfo{pages}{063608} (\bibinfo{year}{2005}).

\bibitem[{\citenamefont{Hutchinson et~al.}(1997)\citenamefont{Hutchinson,
  Zaremba, and Griffin}}]{Hutchinson1997a}
\bibinfo{author}{\bibfnamefont{D.~A.~W.}~\bibnamefont{Hutchinson}},
  \bibinfo{author}{\bibfnamefont{E.}~\bibnamefont{Zaremba}}, \bibnamefont{and}
  \bibinfo{author}{\bibfnamefont{A.}~\bibnamefont{Griffin}},
  \bibinfo{journal}{Phys. Rev. Lett.} \textbf{\bibinfo{volume}{78}},
  \bibinfo{pages}{1842} (\bibinfo{year}{1997}).

\bibitem[{\citenamefont{Proukakis et~al.}(1998)\citenamefont{Proukakis, Morgan,
  Choi, and Burnett}}]{Proukakis1998a}
\bibinfo{author}{\bibfnamefont{N.~P.} \bibnamefont{Proukakis}},
  \bibinfo{author}{\bibfnamefont{S.~A.} \bibnamefont{Morgan}},
  \bibinfo{author}{\bibfnamefont{S.}~\bibnamefont{Choi}}, \bibnamefont{and}
  \bibinfo{author}{\bibfnamefont{K.}~\bibnamefont{Burnett}},
  \bibinfo{journal}{Phys. Rev. A} \textbf{\bibinfo{volume}{58}},
  \bibinfo{pages}{2435} (\bibinfo{year}{1998}).

\bibitem[{\citenamefont{Gerbier
  et~al.}(2004{\natexlab{b}})\citenamefont{Gerbier, Thywissen, Richard,
  Hugbart, Bouyer, and Aspect}}]{Gerbier2004a}
\bibinfo{author}{\bibfnamefont{F.}~\bibnamefont{Gerbier}},
  \bibinfo{author}{\bibfnamefont{J.~H.} \bibnamefont{Thywissen}},
  \bibinfo{author}{\bibfnamefont{S.}~\bibnamefont{Richard}},
  \bibinfo{author}{\bibfnamefont{M.}~\bibnamefont{Hugbart}},
  \bibinfo{author}{\bibfnamefont{P.}~\bibnamefont{Bouyer}}, \bibnamefont{and}
  \bibinfo{author}{\bibfnamefont{A.}~\bibnamefont{Aspect}},
  \bibinfo{journal}{Phys. Rev. Lett.} \textbf{\bibinfo{volume}{92}},
  \bibinfo{pages}{030405} (\bibinfo{year}{2004}{\natexlab{b}}).

\end{thebibliography}
\end{document}